# Resilience-Oriented Operation of Micro-Grids in both Grid-Connected and Isolated Conditions within Sustainable Active Distribution Networks

**Saeed Behzadi, Amir Bagheri***, **Abbas Rabiee**

**Department of Electrical Engineering, University of Zanjan, Zanjan, Iran**

* Corresponding Author (Email: a.bagheri@znu.ac.ir)

**Abstract**

Due to the increasing occurrence of natural disasters, importance of maintaining sustainable energy for cities and society is felt more than ever. On the other hand, power loss reduction is a challenging issue of active distribution networks (ADNs). Therefore, the distribution network operators (DNOs) should have a certain view on these two problems in today's smart grids. In this paper, a new convex optimization model is proposed with two objective functions including energy loss reduction in normal operating mode and system load shedding minimization in critical conditions after the occurrence of natural disasters. This purpose is fulfilled through optimal allocation of distributed generation (DG) units from both conventional and renewable types as well as energy storage systems (ESSs). In addition, a new formulation has been derived to form optimal micro-grids (MGs) aiming at energy loss reduction in normal operating condition and resiliency index improvement under emergency situations. The developed model is implemented in GAMS software and the studies have been tested and analyzed on the IEEE 33-bus system. The results verify the effectiveness of the proposed method in terms of energy loss reduction as well as resilience enhancement in extreme operation condition following severe disruptions in the system.

**Keywords:** ADN, Resiliency, Micro-grid formation, Reconfiguration, Distributed energy resources.

**Nomenclature**

    **Indices**
    $i, j$                        Buses
    $T$                          Time
    $m$                        Micro-grids
    $c$                          System operating mode ($c_1$: grid connected mode, $c_2$: isolated mode)
    $\ell$                          Line between buses $i$ and $j$
    **Sets**
    $\Omega_N$                    Set of buses
    $\Omega_L$                    Set of branches

| Symbol | Description |
|---|---|
| $\Omega_T$ | Set of time |
| $\Omega_C$ | Set of system operating mode |
| $\Omega_S$ | Set of slack buses |
| $\Omega_{MG}$ | Set of micro-grids |
| $\Omega_{DG}$ | Set of candidate buses for installing synchronous generators |
| $\Omega_{ESS}$ | Set of candidate buses for installing energy storage systems |
| $\Omega_{PV}$ | Set of candidate buses for installing photovoltaic units |
| $\Omega_{WT}$ | Set of candidate buses for installing wind turbines |

**Parameters**

| Symbol | Description |
|---|---|
| $w_1 / w_2$ | Weighting coefficient for the objective functions |
| $R_\ell^{Line}$ | Line resistance |
| $X_\ell^{Line}$ | Line reactance |
| $f_i$ | Load priority factor |
| $P_i^L / Q_i^L$ | Peak value of active/reactive load demand |
| $\lambda_t$ | Daily load factor |
| $PF_c$ | Generation power factor of synchronous DGs |
| $bigM$ | A big number |
| $A_{\ell i} / B_{\ell i}$ | $\ell i - th$ element of the bus-line matrix |
| $V_{Min} / V_{Max}$ | Minimum/maximum voltage of buses |
| $I_\ell^{Max}$ | Maximum capacity of lines (in Ampere) |
| $P_{Max}^G / Q_{Max}^G$ | Maximum active/reactive power injected into the system from the upstream network |
| $P_{Min}^{net} / P_{Max}^{net}$ | Minimum/maximum active power passing through lines |
| $Q_{Min}^{net} / Q_{Max}^{net}$ | Minimum/maximum reactive power passing through lines |
| $flow_{Min}^a / flow_{Max}^a$ | Minimum/maximum hypothetical active power that can be passed through lines in mode $c_1$ |
| $flow_{Min}^b / flow_{Max}^b$ | Minimum/maximum hypothetical active power that can be passed through network lines in mode $c_2$ |
| $Gen_{Max}^a$ | Maximum hypothetical active power that can be injected in mode $c_1$ |
| $Gen_{Max}^b$ | Maximum hypothetical active power that can be injected in mode $c_2$ |
| $S_{Max}^{DG}$ | Capacity of synchronous generators |
| $DG^{Max}$ | Maximum number of synchronous generators that can be installed in the network |
| $SOC^{Max}$ | Maximum energy storage capacity of ESS units |
| $P_{Max}^{ESS}$ | Maximum active power of ESS units |
| $EFC / EFD$ | Charge/discharge efficiency of ESS units |
| $V_t^{wind}$ | Wind speed |
| $P_{rated}$ | Rated power of wind turbines |
| $P_{i,c,t}^{WT,capacity}$ | Maximum active power capacity of wind generators |
| $P_{i,c,t}^{PV,capacity}$ | Maximum power capacity of PV unit |
| $V_{c_{in}} / V_{c_{out}}$ | Cut-in/cut-out speeds of wind turbine |
| $V_{rated}$ | Rated speed of wind turbine |
| $si_t$ | Solar radiation intensity |
| $I_{SC}^{PV}$ | Short-circuit current of PV |
| $V_{OC}^{PV}$ | Open-circuit voltage of PV |
| $K_I$ | Temperature coefficient of current in PV unit |
| $K_V$ | Temperature coefficient of voltage in PV unit |
| $T_t^A$ | Ambient temperature |
| $N_{OT}$ | Nominal operating temperature in PV unit |
| $V_{MPP}$ | Maximum power point tracking voltage in PV unit |
| $I_{MPP}$ | Maximum power point tracking current in PV unit |
| $T_t^{cg}$ | Temperature of PV unit |
| $I_t^g$ | Output current of PV unit |
| $V_t^g$ | Output voltage of PV unit |
| $FF$ | Filling factor of PV unit |
| $N_{PV}$ | Number of PV cells |

**Variables**

| | |
|---|---|
| $OF$ | Main objective function |
| $OF_1$, $OF_2$ | Objective function for operating modes of $c_1$ and $c_2$ |
| $OF_1^{optimum}$ | Optimal value of the first objective function without the influence of second objective function |
| $OF_2^{optimum}$ | Optimal value of the second objective function without the influence of first objective function |
| $RI$ | System resilience index |
| $V_{i,c,t}$ | Bus voltage magnitude |
| $I_{\ell,c,t}$ | Line current |
| $U_{i,c,t}$ | Square of bus voltage |
| $J_{\ell,c,t}$ | Square of line current |
| $P_{i,c,t}^{Lsh} / Q_{i,c,t}^{Lsh}$ | Active/reactive power of load shedding |
| $P_{i,c,t,m}^{Lsh,MG} / Q_{i,c,t,m}^{Lsh,MG}$ | Active/reactive power of load shedding in micro-grid |
| $P_{i,c,t,m}^{Lsh,island}$ | Hypothetical active power of load shedding in micro-grids |
| $P^{L_{total}}$ | Total system load |
| $P_c^{Lsh_{total}}$ | Total load shedding |
| $P_{i,j,c,t}^{net} / Q_{i,j,c,t}^{net}$ | Active/reactive power passing through the lines |
| $P_{i,c,t}^G / Q_{i,c,t}^G$ | Active/reactive power injected into the system from the upstream network |
| $P_{i,c,t}^{DG} / Q_{i,c,t}^{DG}$ | Active/reactive power injected into the system by synchronous generators |
| $P_{i,c,t}^{WT}$ | Active power injected into the system by wind generators |
| $P_{i,c,t}^{PV}$ | Active power injected into the system by PVs |
| $X_{i,j,c}$ | Binary variable of connection/disconnection of lines |
| $X_{i,j,c,m}^{MG}$ | Binary variable of connection/disconnection of lines in micro-grid |
| $a_{i,c,m}$ | Binary variable indicating the presence of bus $i$ in micro-grid $m$ |
| $r_{i,j,m}^{pos} / r_{i,j,m}^{neg}$ | Positive/negative binary variable used for linearization of absolute function |
| $DG_i$ | Binary variable to allocate synchronous generators |
| $DG_{i,c,m}^{MG}$ | Binary variable indicating the presence of DG *i* in micro-grid *m* |
| $DG_{i,c,m}^{island}$ | Binary variable for the presence of at least one synchronous generator in each micro-grid |
| $ich_{i,t} / idch_{i,t}$ | Charge/discharge binary variable of ESS |
| $Gen_{i,c,t}^{a}$ | Hypothetical injected active power in mode $c_1$ |
| $Gen_{i,c,t,m}^{b}$ | Hypothetical injected active power in mode $c_2$ |
| $flow_{i,j,c,t}^{a}$ | Hypothetical active power passing through lines in mode $c_1$ |
| $flow_{i,j,c,t,m}^{b}$ | Hypothetical active power passing through lines in mode $c_2$ |
| $SOC_{i,c,t}$ | State of charge for ESS units |
| $PC_{i,c,t} / PD_{i,c,t}$ | Active charging/discharging capacity of ESS units |

## 1. Introduction

In recent years, due to climate change, natural disasters such as floods, hurricanes, earthquakes, ice storms, dust storms, etc. have been intensified in many countries [1]. Modern and sustainable distribution systems have the ability to maintain network security and proper feeding of loads in the face of daily weather challenges such as continuous rain, snow and wind. However, some extreme weather events with a low probability of occurrence can still cause long-term power outages in distribution network (DN) leading to numerous problems. Events like Sandy hurricane, which left 7.5 million customers without energy across Washington and 15 other states, expresses the importance and urgency of improving power system resiliency [2]. The concept of "power system resilience" is a measure to evaluate the ability of a system to withstand against significant power outages caused by natural disasters or intentional attacks, and loads restoration after the occurrence of these types of incidents [1-3]. Prevention, survivability, and recovery are three main parts of the power system resiliency [4]. Different approaches can be employed to improve distribution systems resilience in front of natural disasters. The main traditional way of resiliency improvement is the infrastructure reinforcement such as construction of additional lines/transformers [5], enhancing towers [6], etc. However, these approaches require huge values of investment costs, and they are faced with long implementation periods as well as right-of-way and environmental concerns. Nowadays, small-scale energy resources including DGs and energy storage units are widely integrated into distribution networks because of their numerous benefits. The DG units can be divided into synchronous generators and renewable-energy-based ones such as photovoltaics (PVs) and wind turbines (WTs). The benefits of DGs and ESS units include lower environmental emissions, fast installation time, improving voltage profile, power loss reduction, and assisting in reliability enhancement [7, 8]. In this regard, DG and ESS units can be utilized also to improve distribution networks resiliency. For this aim, these units should be optimally placed within the system [9]. Another low-cost and more accessible alternative for the resiliency improvement is formation of micro-grids after occurrence of disaster and disconnection of some parts of the network. This action can be fulfilled by the aid of opening/closing tie lines and switches regarding system constraints [10]. Although the resiliency improvement is highly important for today's networks, it should be noted that the system is in normal operating conditions at most of the times. On this basis, distribution network operators have specific insight to optimal operation of the system during normal operation. The main operational parameter of the distribution system is the power loss which the DNOs always try to reduce it by means of different methods [11]. DG and ESS allocation are among the alternatives that help to power loss reduction in DNs along with their vital role in resiliency improvement [12]. On the other hand, opening/closing of tie lines and switches is the other solution of power loss reduction. This action is named as "Reconfiguration" in normal operating condition [13]. Regarding the above discussion, resiliency improvement of distribution systems in emergency conditions and power loss reduction under normal operating condition are from are important priorities of DNOs. In the following the recent works in this area are reviewed to investigate their features and shortcomings.

Ref. [14] has addressed different types of resilience assessment techniques, along with a comparison based on several criteria, including resilience assessment algorithm, graph, methodology, and equations. Also, the impact and influence of the extreme events on the power system and the role of DG in solving the challenges caused by these events have been highlighted.

In [15, 16], the effect of distributed energy resources (DERs) such as renewable energy sources, energy storage systems, demand response, reconfiguration, and on-load tap changers (OLTCs) has been evaluated on minimizing the energy procurement costs on an hourly basis. Using line flow based (LFB) power flow equations and other convexifications, the main model is transformed to a convex mixed-integer second order conic programming (MISOCP) model. The simulation results demonstrate significant impact of DERs in reducing energy procurement cost as well as improving voltage profile. This paper dedicates to optimizing normal system operation using DER units without considering emergency conditions.

A bi-level optimization model is presented in [17] for reconfiguration of distribution system for improving the network resiliency against severe weather events such as storm and hurricane aiming at minimization of load outage cost. To fulfill this goal, at first, the vulnerability of distribution network poles is evaluated by a model to estimate the damages imposed by the threat. Then, a network reconfiguration strategy is employed to minimize the expected cost of load outage based on forecasting of possible failed lines and predicted wind speed before the storm. This paper has not regarded network's normal operation. Moreover, the optimization model is mixed-integer nonlinear programming (MINLP) which is non-convex.

Ref. [18] deals with a new model based on mixed integer linear programming (MILP) for modeling and evaluating resiliency of smart distribution systems. In addition, using a two-stage framework based on stochastic programming, effect of increasing the share of renewable energy sources in the network and their related uncertainties on the system resiliency is investigated. In [19], a resilience-based critical load restoration method using micro-grids formation after natural disasters is proposed in which the load recovery problem is modeled as a constrained stochastic program, and unbalanced three-phase power flow is used in an MILP model. In [20], a resilience-based operational approach using MILP is proposed in distribution systems having radial structure to restore critical loads after natural disaster occurrence. The restoration is accomplished by forming several micro-grids supplied by DGs. This paper has not placed DG units, and their location is considered fixed within the network. In [21], a new micro-grid formation model considering the power losses and voltage constraints is presented as an emergency operation strategy when extreme weather conditions occur. The linearized form of DistFlow [22] formulation as the power flow analysis is employed to convert the model to an MILP one. A more general advantage of the proposed method is formation of micro-grids with lower power losses than traditional methods, and this means that more loads can be recovered. This paper has not addressed renewable-energy-based DG units. In [23], a method is proposed in which after the optimal allocation of distributed generation resources, the system is divided into several micro-grids to reduce the amount of load outages in large-scale events. A multi-objective optimization with two goals including total load shedding as well as total voltage deviation has been formulated as a MINLP model, and a heuristic approach named exchange market algorithm (EMA) is developed to solve the problem. This work has not regarded renewable DGs and energy storage units. In order to improve resiliency of distribution systems, an integrated framework is developed in [24] in which two models, called defender–attacker–defender, are made for finding proper solution to reduce load shedding during sever events. These models are carried out in three levels. Hardening of system equipment, calculating the highest load shedding, and reconfiguration are the actions implemented in the first model. In the second model, reinforcement plan and the worst-case attack constitute the first and second levels, and the optimal DG placement for supplying the islanded micro-grids compose the third level. This paper's studies are devoted to only the peak hour of the system, and the load variation during different hours has not been taken into account.

Moreover, this paper has considered only the radial structure in the islanded mode of operation. Also, the DG allocation in this paper is implemented with regarding the emergency condition, and normal operation of the network does not play any role in the DG placement procedure. Ref. [25] presents a MILP model for resilience-oriented operation of micro-grids in active distribution networks using a multi-objective stochastic modeling. The probabilistic behavior of disaster has been modeled using its probability distribution function in the presence of renewable power generations and electric vehicles as transportable storages. Both normal and emergency operating modes have been considered to improve the performance of distribution system in terms of operating costs and meeting technical constraints, and also, minimizing the load shedding during emergency conditions. The configuration of DN is fixed in this study, and there is no possibility of reconfiguration to optimize DN's performance under normal and emergency conditions. Also, the DG units' locations are fixed and they are not optimally placed. To avoid load shedding after natural disasters, a two-stage optimization model is presented in [26] for optimizing investment in mobile energy storage units and determining their routing to form dynamic MGs. By applying the progressive hedging algorithm, the proposed problem is modeled as MISOCP model which is solved using off-the-shelf solvers. This paper has not allocated DG units, and their locations are fixed. After reviewing the related literature, their characteristic can be summarized as Table 1.

Table 1: Comparison of recent literature with the proposed model

| | Master Unit Placement | WT | ESS | PV | Master And Slave Units | Micro Grid Formation | Reconfiguration | Radiality Constraint | Mesh Constraint | Resiliency Index | LFB | Power Loss | Load Shedding |
|---|---|---|---|---|---|---|---|---|---|---|---|---|---|
| [1] (2020) | | ✓ | ✓ | | ✓ | ✓ | ✓ | ✓ | | ✓ | | | |
| [2] (2020) | | | ✓ | | ✓ | ✓ | ✓ | ✓ | | | | | |
| [15] (2021) | | ✓ | ✓ | ✓ | | | ✓ | ✓ | | | ✓ | | |
| [16] (2019) | | ✓ | ✓ | ✓ | | | ✓ | ✓ | | | ✓ | ✓ | |
| [17] (2019) | | | | | | | ✓ | ✓ | | ✓ | | | ✓ |
| [23] (2021) | ✓ | | | | | ✓ | ✓ | ✓ | | | | | ✓ |
| [24] (2020) | | ✓ | ✓ | | | ✓ | ✓ | ✓ | | ✓ | | | ✓ |
| [25] (2022) | | ✓ | ✓ | ✓ | ✓ | | | ✓ | | ✓ | | | ✓ |
| Proposed model | ✓ | ✓ | ✓ | ✓ | ✓ | ✓ | ✓ | ✓ | ✓ | ✓ | ✓ | ✓ | ✓ |

This paper proposes a new optimization model with two objective functions. The first objective function includes energy loss reduction in normal operating condition, and the second one consists of system load shedding minimization in critical conditions after occurrence of natural disasters. These two objectives have been simultaneously considered within the optimization process. These goals are attained through optimal allocation of DG units from both conventional and renewable types as well as ESS resources. In addition, a new formulation has been derived to optimal reconfiguration aiming at energy loss reduction in normal operating condition, and optimal micro-grids formation for the aim of resiliency index improvement under emergency conditions. By using new formulations, all the required relations have been convexified to form a mixed-integer quadratically-constrained programming (MIQCP). Moreover, the line flow based (LFB) algorithm is utilized for the AC power flow calculations. Furthermore, comparison of radial and mesh structures in micro-grids after the fault occurrence have been made. In addition, in the emergency conditions, the results have been compared for different number of MGs. The obtained results demonstrate effectiveness of the proposed model. The contributions of the submitted paper can be summarized as follows:

- Proposing a resilience-oriented operation of micro-grids;
- Both grid-connected and isolated conditions have been considered;
- New formulations have been developed for reconfiguration and micro-grid formation;
- New formulations have been extended for DG allocation in the presence of DERs;
- Convexifying the relations along with LFB model of AC power flow.

## 2. Problem Formulations

### 2.1. Objective Function

In distribution networks, the power loss is an important criterion in optimal operation under normal operating conditions. Also, regarding natural disasters and critical conditions, the purpose of the network operator is to minimize network load shedding. Therefore, the objective function proposed in this paper consists of two terms as (1). The first term ($OF_1$) is distribution system energy losses according to (2), and the second one ($OF_2$) includes total load shedding after the fault occurrence which is represented by (3). In the formulas, the states $c_1$ and $c_2$ express the normal and emergency modes of operation, respectively. The two objectives have been combined with each other by using weighting factors of $w_1$ and $w_2$. In addition, the objectives have been normalized by their individual optimum values. In the main objective function of (1), the value of $OF_1^{optimum}$ is obtained by optimization of the $OF_1$ without considering $OF_2$. Also, $OF_2^{optimum}$ is gained by optimization of $OF_2$ without regarding $OF_1$. If the optimum value of $OF_1$ and $OF_2$ are not used, it would be needed to consider many weighting factors of $w_1$ and $w_2$ to evaluate the impact of the first and second objective functions. To avoid using several weighting factors, the normalization method has been employed in this paper such that the number of required $w_1$ and $w_2$ is minimum. In this paper, it has been considered different priorities for the loads priorities which are shown by $f_i$ in (3). This is to apply the importance of loads in the process of load restoration in emergency conditions.

$$OF = Min\left\{\frac{\left(\frac{w_1 \, OF_1}{OF_1^{optimum}} + \frac{w_2 \, OF_2}{OF_2^{optimum}}\right)}{w_1 + w_2}\right\} \quad (1)$$

$$OF_1 = \sum_{t \in \Omega_T} \sum_{\ell \in \Omega_L} R_\ell^{Line} J_{\ell,c,t} \qquad ; \forall c \in c_1 \quad (2)$$

$$OF_2 = \sum_{t \in \Omega_T} \sum_{i \in \Omega_N} f_i P_{i,c,t}^{Lsh} \qquad ; \forall c \in c_2 \quad (3)$$

$$\frac{w_1 + w_2}{4} = 1 \qquad ; \forall \; 0 \leq w_1 \leq 4 \quad (4)$$

## 2.2. Problem constraints

The proposed problem is subject to different constraints which are described subsequently.

### 2.2.1. Power flow equations

In distribution systems, the AC power flow relations are required in order to accurately calculate power loss as well as voltage magnitude of buses. The conventional AC power flow equations are non-linear and non-convex which increases execution time and makes complex the optimization procedure. Therefore, in this paper, the LFB model [15], which is a convexified version of AC power flow, has been employed as relations (5)-(13).

$\forall i, j \in \Omega_N, \forall t \in \Omega_T, \forall \ell \in \Omega_L, \forall c \in \Omega_C :$

$$\sum_{j \in \Omega_N} A_{\ell i} P_{i,j,c,t}^{net} = P_{i,c,t}^G + P_{i,c,t}^{DG} + P_{i,c,t}^{WT} + P_{i,c,t}^{PV} + P_{i,c,t}^{Lsh} + PD_{i,c,t} - PC_{i,c,t} - \lambda_t P_i^L - \sum_{j \in \Omega_N} B_{\ell i} R_\ell^{Line} J_{\ell,c,t} \quad (5)$$

$$\sum_{j \in \Omega_N} A_{\ell i} Q_{i,j,c,t}^{net} = Q_{i,c,t}^G + Q_{i,c,t}^{DG} + Q_{i,c,t}^{Lsh} - \lambda_t Q_i^L - \sum_{j \in \Omega_N} B_{\ell i} X_\ell^{Line} J_{\ell,c,t} \quad (6)$$

$$-(1 - X_{i,j,c})bigM \leq U_{j,c,t} + 2\left(R_\ell^{Line} P_{i,j,c,t}^{net} + X_\ell^{Line} Q_{i,j,c,t}^{net}\right) - U_{i,c,t} + \left[\left(R_\ell^{Line}\right)^2 + \left(X_\ell^{Line}\right)^2\right] J_{\ell,c,t}$$
$$\leq (1 - X_{i,j,c})bigM \quad (7)$$

$$\left(P_{i,j,c,t}^{net}\right)^2 + \left(Q_{i,j,c,t}^{net}\right)^2 = J_{\ell,c,t} U_{j,c,t} \quad (8)$$

$$P_{i,j,c,t}^{net} = -P_{j,i,c,t}^{net} \quad (9)$$

$$Q_{i,j,c,t}^{net} = -Q_{j,i,c,t}^{net} \quad (10)$$

$$-X_{i,j,c} P_{Min}^{net} \leq P_{i,j,c,t}^{net} \leq X_{i,j,c} P_{Max}^{net} \quad (11)$$

$$-X_{i,j,c} Q_{Min}^{net} \leq Q_{i,j,c,t}^{net} \leq X_{i,j,c} Q_{Max}^{net} \quad (12)$$

$$X_{i,j,c} = X_{j,i,c} \quad (13)$$

In (5) and (6), $A_{\ell i}$ is the $\ell i - th$ element of the bus-line matrix. It is equal to 1 if bus $i$ is the sending bus of line $\ell$. If bus $i$ is the receiving bus of line $\ell$, $A_{\ell i}$ equals to -1. Otherwise, it will be zero. Also, $B_{\ell i}$ is the same as $A_{\ell i}$ by replacing 1 with 0. $P_{i,j,c,t}^{net}$ and $Q_{i,j,c,t}^{net}$ are the active and reactive power flows through line $\ell$ in mode $c$, at time $t$, respectively. $P_{i,c,t}^{G}$ and $Q_{i,c,t}^{G}$ are active and reactive power injected into the network from the upstream system. $P_{i,c,t}^{DG}$ and $Q_{i,c,t}^{DG}$ are the active and reactive power injected into the system by master DGs. $P_{i,c,t}^{WT}$ and $P_{i,c,t}^{PV}$ are the active power generated by wind generators and PVs. Also, $PC_{i,c,t}$ and $PD_{i,c,t}$ denote charge and discharge powers of ESS units. In this paper, the synchronous generators have been considered as master DGs, and wind generators and PV units are considered as slave ones. It should be noticed that in the emergency conditions where distribution network is isolated from the main supplying substation, there must be at least one master DG within each islanded part. In (7), relation of voltages of two nearby buses has been defined, where $X_{i,j,c}$ is a binary variable indicating that the line between buses $i$ and $j$ in operating mode of $c$ is connected or not. Eq. (8) is the nodal relationship between power, voltage and current. In (9) and (10), the relation between active and reactive powers of sending and receiving buses can be seen. It should be noted that $P_{j,i,c,t}^{net}$ and $Q_{j,i,c,t}^{net}$ are active and reactive powers sent by node $j$ toward node $i$ at the side of node $i$. That is, the power loss of line $i$-$j$ has not been considered in (9) and (10). Relations (11) and (12) guarantee the power passing through the disconnected lines to be zero. Eq. (13) ensures two-way connection of the lines between two ending buses. The constraints related to buses' voltages and branches' currents have been given in (14)-(18). These constraints must be satisfied for both normal and emergency conditions.

$$\begin{cases} V_{i,c,t}=1 & ;\forall c \in c_1, i \in \Omega_S \\ V_{Min} \leq V_{i,c,t} \leq V_{Max} \end{cases} \quad (14)$$

$$U_{i,c,t}=(V_{i,c,t})^2 \quad (15)$$

$$(V_{Min})^2 \leq U_{i,c,t} \leq (V_{Max})^2 \quad (16)$$

$$0 \leq J_{\ell,c,t} \leq X_{i,j,c}(I_\ell^{Max})^2 \quad (17)$$

$$J_{\ell,c,t}=(I_{\ell,c,t})^2 \quad (18)$$

By equation (17), it is guaranteed that the current flow is zero in disconnected lines (when $X_{i,j,c}=0$), and for the connected lines (when $X_{i,j,c}=1$), the current flow must be lower than the thermal capacity of each line. Relations (19) and (20) express limitation of injected power from the upstream network.

$$\begin{cases} 0 \leq P_{i,c,t}^G \leq P_{Max}^G & ;\forall i \in \Omega_S, c \in c_1 \\ P_{i,c,t}^G=0 & ; \quad otherwise \end{cases} \quad (19)$$

$$\begin{cases} 0 \leq Q_{i,c,t}^G \leq Q_{Max}^G & ;\forall i \in \Omega_S, c \in c_1 \\ Q_{i,c,t}^G=0 & ; \quad otherwise \end{cases} \quad (20)$$

Except for equation (8), all the power flow constraints in the LFB model are convex and linear. To convexify (8), the conic relaxation method can be used. For this purpose, equation (8) is rewritten as (21).

$$\left(P_{i,j,c,t}^{net}\right)^2 + \left(Q_{i,j,c,t}^{net}\right)^2 \leq J_{\ell,c,t} U_{j,c,t} \tag{21}$$

Now, the above problem is a mixed-integer second-order conic programming optimization model that can be easily solved with GUROBI, MOSEK, CPLEX, and other appropriate solvers [16].

### 2.2.2. Micro-grid formation constraints

This paper proposes a new formulation for micro-grid formation as an innovation. In this formulation, a binary variable as $a_{i,c,m}$ is defined which indicates the presence of bus $i$ in micro-grid $m$. Based on this formation, all the buses of network should be in a specific island. Therefore, there is no isolated bus within the network. In this paper, the system has the ability to be divided into one integrated island, or two and more separated islands. As the islanding mode is intended only for the post-disaster conditions, islanding is limited to mode $c_2$; this issue has been shown in (22). In this case, the system is disconnected from the upstream network. For this reason, the reference (main substation) bus, shown by $i=1$, should no longer exist in any island; this is stated in relation (23).

$\forall i \in \Omega_N, \forall m \in \Omega_{MG}:$

$$a_{i,c,m} = 0 \quad ; \forall c \in c_1 \tag{22}$$

$$a_{i,c,m} = 0 \quad ; \forall c \in c_2, i = 1 \tag{23}$$

In the following, the relations required for the micro-grid formation are discussed. For this aim, the first issue that should be considered is that each bus and each line in the network belong to only one island (micro-grid), which is expressed by (24)-(27).

$\forall i, j \in \Omega_N, \forall t \in \Omega_T, \forall \ell \in \Omega_L, \forall m \in \Omega_{MG}, \forall c \in c_2:$

$$\sum_{m \in \Omega_{MG}} a_{i,c,m} \leq 1 \tag{24}$$

$$X_{i,j,c,m}^{MG} = X_{j,i,c,m}^{MG} \tag{25}$$

$$\sum_{m \in \Omega_{MG}} X_{i,j,c,m}^{MG} \leq 1 \tag{26}$$

$$\sum_{m \in \Omega_{MG}} X_{i,j,c,m}^{MG} = X_{i,j,c} \tag{27}$$

One of the main limitations of islanding is that there should not be a connecting line between any of the buses of two different islands. This matter is expressed by (28)-(30):

$$\left| a_{i,c,m} - a_{j,c,m} \right| \leq 1 - X_{i,j,c,m}^{MG} \tag{28}$$

$$X_{i,j,c,m}^{MG} \leq a_{i,c,m} \tag{29}$$

$$X_{i,j,c,m}^{MG} \leq a_{j,c,m} \tag{30}$$

Due to presence of absolute function in (28), this relation is non-linear. To linearize (28), relations (31)-(33) are employed in this paper. The parameters related to these equations are defined in the nomenclature.

$$r_{i,j,m}^{pos} + r_{i,j,m}^{neg} \leq 1 - X_{i,j,c,m}^{MG} \tag{31}$$

$$a_{i,c,m} - a_{j,c,m} = r_{i,j,m}^{pos} - r_{i,j,m}^{neg} \tag{32}$$

$$r_{i,j,m}^{pos} + r_{i,j,m}^{neg} \leq 1 \tag{33}$$

### 2.2.3. Constraints related to radial and mesh configurations

To obtain radial structure for the network, the following two conditions must be satisfied:

(a) The number of network lines must be equal to the number of nodes minus one;

(b) There must be a path from the slack bus in the circuit to all buses in the system.

The reason why condition (a) is required is that if condition (b) is met, no loop can be formed in the network. The reason for the requirement of condition (b) is that no bus nor a set of buses is isolated in the network; if this happens (isolation of buses), a loop will be created in the system, which contradicts the radiality condition [27]. To reach a structure without restrictions and without the requirement of being radial (the network can be arranged radially or mesh), condition (b) must be satisfied and only condition (a) should be rewritten as follows:

- The number of network lines is greater than the number of nodes minus one.

In this paper, the authors have improved the method proposed in [27] to establish a radial and mesh structures in micro-grids as another innovation.

*Radiality constraint in normal operating condition*

In the case that the system is connected to the upstream network, the distribution system is operated in radial structure. In the following, the required formulation for satisfying this constraint are given.

$\forall i, j \in \Omega_N, \forall t \in \Omega_T, \forall \ell \in \Omega_L, c \in c_1:$

Condition (a):

$$\sum_{\ell \in \Omega_L} X_{i,j,c} = 2\left[\left(\sum_{i \in \Omega_N} i\right) - 1\right] \tag{34}$$

Condition (b):

$$Gen_{i,c,t}^a - \lambda_t P_i^L = \sum_{j \in \Omega_N} flow_{i,j,c,t}^a \tag{35}$$

$$flow_{i,j,c,t}^a + flow_{j,i,c,t}^a = 0 \tag{36}$$

$$X_{i,j,c} \, flow_{Min}^a \leq flow_{i,j,c,t}^a \leq X_{i,j,c} \, flow_{Max}^a \tag{37}$$

$$\begin{cases} 0 \leq Gen_{i,c,t}^{a} \leq Gen_{Max}^{a} & ; \forall i \in \Omega_S, \forall c \in c_1 \\ Gen_{i,c,t}^{a} = 0 & ; \quad otherwise \end{cases} \quad (38)$$

$$flow_{Min}^{a} \leq flow_{i,j,c,t}^{a} \leq flow_{Max}^{a} \quad (39)$$

In (34), multiplication of 2 is due to the fact that each line is counted twice, once from bus $i$ to $j$ and once from bus $j$ to $i$. By the relations (35)-(39), using the idea of a hypothetical power flow in condition (b), the desired goal can be achieved.

### *Radial and mesh structure constraints in emergency condition*

To implement radial and non-radial structure in the case of critical conditions, the previously stated method should be improved as follows so that the stated conditions are true in each island. Therefore, new relations are presented in this paper to achieve this purpose in micro-grids. In critical conditions, the distribution system has been investigated in two scenarios:

1) First scenario: the distribution system must have a radial structure

2) second scenario: the distribution system can have either radial or mesh structures

- **First scenario**

Relation (40) which is the modified version of (34) satisfies the condition (a) given above.

$\forall i, j \in \Omega_N, \forall t \in \Omega_T, \forall \ell \in \Omega_L, \forall m \in \Omega_{MG}, \forall c \in c_2:$

$$\sum_{\ell \in \Omega_L} X_{i,j,c,m}^{MG} = 2\left[\left(\sum_{i \in \Omega_N} a_{i,c,m}\right) - 1\right] \quad (40)$$

In mode $c_2$, because after the event, the system is divided into separate islands, the second condition (b) must be satisfied in each MG. Hence, the new relations are expressed as follows. Relation (44) guarantees connection of MG buses to each other.

$$Gen_{i,c,t,m}^{b} + P_{i,c,t,m}^{Lsh,island} - \lambda_t a_{i,c,m} P_i^L = \sum_{j \in \Omega_N} flow_{i,j,c,t,m}^{b} \quad (41)$$

$$flow_{i,j,c,t,m}^{b} + flow_{j,i,c,t,m}^{b} = 0 \quad (42)$$

$$X_{i,j,c,m}^{MG} flow_{Min}^{b} \leq flow_{i,j,c,t,m}^{b} \leq X_{i,j,c,m}^{MG} flow_{Max}^{b} \quad (43)$$

$$0 \leq P_{i,c,t,m}^{Lsh,island} \leq 0.99 a_{i,c,m} \lambda_t P_i^L \quad (44)$$

$$\begin{cases} 0 \leq Gen_{i,c,t,m}^{b} \leq Gen_{Max}^{b} DG_{i,c,m}^{island} & ; \forall i \in \Omega_{DG}, \forall c \in c_2 \\ Gen_{i,c,t,m}^{b} = 0 & ; \quad otherwise \end{cases} \quad (45)$$

- **Second scenario**

To set up a mesh structure, to satisfy condition (b) in this scenario, the relations (41)-(45) must be met, and to establish the condition (a), the relation (40) is changed as follows.

$$\sum_{\ell \in \Omega_L} X_{i,j,c,m}^{MG} \geq 2\left[\left(\sum_{i \in \Omega_N} a_{i,c,m}\right) - 1\right] \tag{46}$$

### 2.2.4. Load shedding constraints

It should be noticed that the load shedding is permitted only in the emergency condition. The constraints related to the load shedding in each micro-grid are as (47)-(52). Eq. (47) states the relationship between active and reactive powers. Relations (48) and (49) limit the load shedding to the load of the considered bus.

$\forall i \in \Omega_N, \forall t \in \Omega_T, \forall m \in \Omega_{MG}, \forall c \in c_2:$

$$Q_{i,c,t}^{Lsh} = \left(\frac{\lambda_t Q_i^L}{\lambda_t P_i^L}\right) P_{i,c,t}^{Lsh} \tag{47}$$

$$0 \leq P_{i,c,t,m}^{Lsh,MG} \leq a_{i,c,m} \lambda_t P_i^L \tag{48}$$

$$0 \leq Q_{i,c,t,m}^{Lsh,MG} \leq a_{i,c,m} \lambda_t Q_i^L \tag{49}$$

$$P_{i,c,t}^{Lsh} = \sum_{m \in \Omega_{MG}} P_{i,c,t,m}^{Lsh,MG} \tag{50}$$

$$Q_{i,c,t}^{Lsh} = \sum_{m \in \Omega_{MG}} Q_{i,c,t,m}^{Lsh,MG} \tag{51}$$

$$\begin{cases} P_{i,c,t}^{Lsh} = 0 & \forall c \in c_1 \\ Q_{i,c,t}^{Lsh} = 0 & \forall c \in c_1 \end{cases} \tag{52}$$

### 2.2.5. Constraints of synchronous DGs

Relations (53) and (54) show active and reactive power generation limits of DGs. In (54), $DG_i$ is a binary variable indicating installation of DG on bus $i$, and $S_{Max}^{DG}$ denotes maximum capacity of DG unit. Also, in relation (55), $DG^{Max}$ is maximum number of synchronous generators.

$\forall i \in \Omega_{DG}, \forall t \in \Omega_T, \forall m \in \Omega_{MG}, \forall c \in \Omega_C:$

$$-\tan\left(\cos^{-1}(PF_c)\right) P_{i,c,t}^{DG} \leq Q_{i,c,t}^{DG} \leq \tan\left(\cos^{-1}(PF_c)\right) P_{i,c,t}^{DG} \tag{53}$$

$$\left(P_{i,c,t}^{DG}\right)^2 + \left(Q_{i,c,t}^{DG}\right)^2 \leq DG_i \left(S_{Max}^{DG}\right)^2 \tag{54}$$

$$\sum_{i \in \Omega_N} DG_i \leq DG^{Max} \tag{55}$$

In this paper, placement of master DGs has been implemented using a new formulation regarding the MG formation as (56)-(63). These relations ensure presence of at least one master DG within each micro-grid for the aim of guaranteeing frequency stability. It is worth mentioning that the relations have been extracted in a way that the convexity of the model is fulfilled.

$$\sum_{m \in \Omega_{MG}} DG_{i,c,m}^{MG} \leq DG^{Max} - \left(\sum_{m \in \Omega_{MG}} m\right) + 1 \quad ; \forall c \in c_2 \tag{56}$$

$$\sum_{i \in \Omega_{DG}} DG_{i,c,m}^{MG} \geq 1 \qquad ;\forall c \in c_2 \tag{57}$$

$$DG_i = \sum_{m \in \Omega_{MG}} DG_{i,c,m}^{MG} \qquad ;\forall c \in c_2 \tag{58}$$

$$\sum_{m \in \Omega_{MG}} \sum_{i \in \Omega_{DG}} DG_{i,c,m}^{island} = \sum_{m \in \Omega_{MG}} m \qquad ;\forall c \in c_2 \tag{59}$$

$$\sum_{i \in \Omega_{DG}} DG_{i,c,m}^{island} = 1 \qquad ;\forall c \in c_2 \tag{60}$$

$$DG_{i,c,m}^{island} \leq DG_{i,c,m}^{MG} \qquad ;\forall c \in c_2 \tag{61}$$

$$\begin{cases} DG_{i,c,m}^{island} = 0 & ;\forall c \in c_1 \\ DG_{i,c,m}^{MG} = 0 & ;\forall c \in c_1 \end{cases} \tag{62}$$

$$a_{i,c,m} \geq DG_{i,c,m}^{MG} \qquad ;\forall c \in c_2 \tag{63}$$

### 2.2.6. Constraints of ESS, PV, and WT units

The equations and limitations related to ESS, WT, and PV units are expressed as follows. Eq. (64) shows the state of charge (energy) of ESS unit at different hours which is constrained by (65). Relations (66) and (67) represent charge/discharge power of each ESS. According to (68), charging and discharging of storage devices cannot be occurred simultaneously. Also, (69) confines the energy of last hour to that of its initial value.

$$\forall i \in \Omega_{ESS}, \forall t \in \Omega_T, \forall c \in \Omega_C:$$

$$SOC_{i,c,t} = SOC_{i,c,(t-1)} + (PC_{i,c,t} EFC) - \left(\frac{PD_{i,c,t}}{EFD}\right) \tag{64}$$

$$0 < SOC_{i,c,t} < SOC^{Max} \tag{65}$$

$$0 \leq PC_{i,c,t} \leq ich_{i,t} P_{Max}^{ESS} \tag{66}$$

$$0 \leq PD_{i,c,t} \leq idch_{i,t} P_{Max}^{ESS} \tag{67}$$

$$ich_{i,t} + idch_{i,t} \leq 1 \tag{68}$$

$$SOC_{i,c,t_{24}} = SOC_{i,c,t_0} \tag{69}$$

For the wind turbine units, relations (70) and (71) are established, where (70) shows generated power of WT according to power-speed characteristics of the wind turbine. Also, (71) represents generation limit of the WT.

$$\forall i \in \Omega_{WT}, \forall t \in \Omega_T, \forall c \in \Omega_C:$$

$$P_{i,c,t}^{WT,capacity} = \begin{cases} 0 & ;\forall V_t^{wind} < V_{c_{in}} \\ P_{rated}\left(\dfrac{V_t^{wind} - V_{c_{in}}}{V_{rated} - V_{c_{in}}}\right) & ;\forall V_{c_{in}} \leq V_t^{wind} < V_{rated} \\ P_{rated} & ;\forall V_{rated} \leq V_t^{wind} < V_{c_{out}} \\ 0 & ;\forall V_t^{wind} \geq V_{c_{out}} \end{cases} \tag{70}$$

$$0 \leq P_{i,c,t}^{WT} \leq P_{i,c,t}^{WT,capacity} \tag{71}$$

The relations governing PV units can be stated as (72)-(77). The generated power of photovoltaic unit is limited by (72) in which $P_{i,c,t}^{PV,capacity}$ shows the available capacity of PV at each hour.

$$\forall i \in \Omega_{PV}, \forall t \in \Omega_T, \forall c \in \Omega_C:$$
$$0 \leq P_{i,c,t}^{PV} \leq P_{i,c,t}^{PV,capacity} \tag{72}$$
$$P_{i,c,t}^{PV,capacity} = N_{PV} V_t^g I_t^g FF \tag{73}$$
$$I_t^g = si_t \left( I_{SC}^{PV} + K_I \left( T_t^{cg} - 25 \right) \right) \tag{74}$$
$$V_t^g = V_{OC}^{PV} - K_V T_t^{cg} \tag{75}$$
$$T_t^{cg} = T_t^A + si_t \left( \frac{N_{OT} - 20}{0.8} \right) \tag{76}$$
$$FF = \frac{V_{MPP} I_{MPP}}{V_{OC}^{PV} I_{SC}^{PV}} \tag{77}$$

### 2.2.7. Resiliency index

The resiliency index in the proposed problem is defined as (78) which shows the difference of total load and total load shedding during the whole time period of the day [1, 24]. In the other words, *RI* denotes the percent of energy not supplied during the emergency condition.

$$RI = \left( \frac{P^{L_{total}} - P_c^{Lsh_{total}}}{P^{L_{total}}} \right) \times 100 \qquad ; \forall c \in c_2 \tag{78}$$
$$P_c^{Lsh_{total}} = \sum_{t \in \Omega_T} \sum_{i \in \Omega_N} P_{i,c,t}^{Lsh} \qquad ; \forall c \in c_2 \tag{79}$$
$$P^{L_{total}} = \sum_{t \in \Omega_T} \lambda_t \sum_{i \in \Omega_N} P_i^L \tag{80}$$

## 3. Numerical study

### 3.1. Test system and input parameters

The proposed model in this paper has been applied to the IEEE 33-bus system as shown in Fig. 1. This system is a 12.66 kV one having 32 main line and 5 tie lines. Twelve number of lines are equipped with sectionalizer as Fig. 1. The total load of this network is 3.715 MW and 2.3 MVAr at the peak load hour. The data of lines and loads have been given in [28]. The load profile during the day is shown in Fig. 2. Also, the load priority factors ($f_i$) for the buses have been shown in Fig. 3. In addition, Figs. 4, 5, and 6 illustrate the wind speed, sun radiation, and ambient temperature during the day, respectively. These data are based on Ref. [29]. In Table 2, the required constant parameters of network equipment have been given [30]. All the simulations have been implemented in the GAMS using the CPLEX solver.

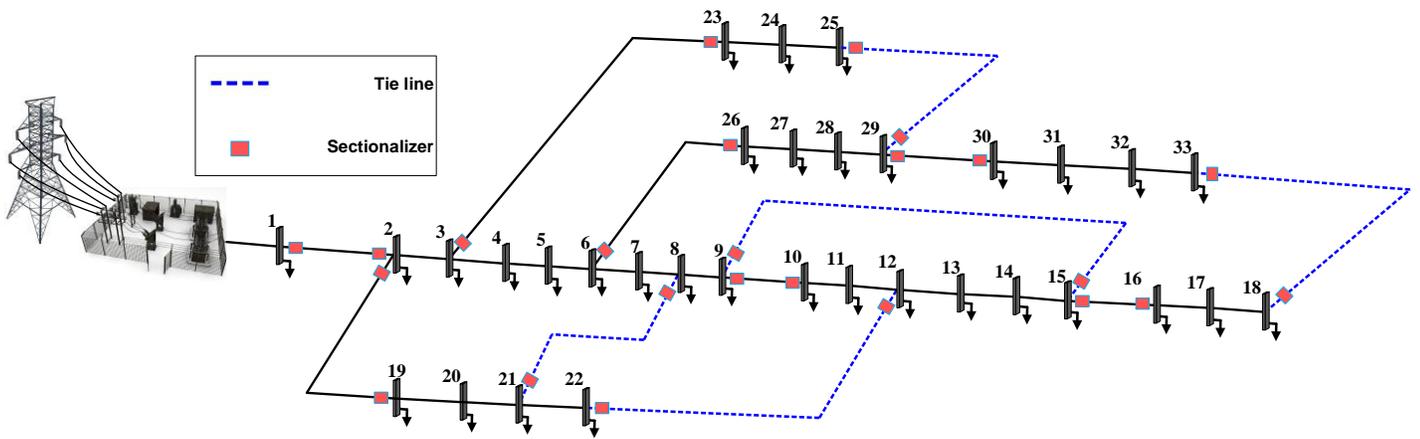

Fig. 1: Single-line diagram of IEEE 33-bus network

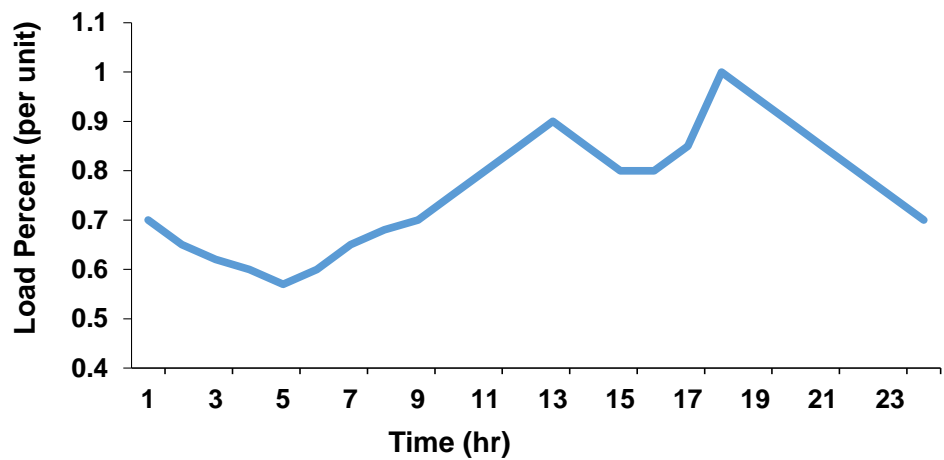

Fig. 2: Daily load profile of the network

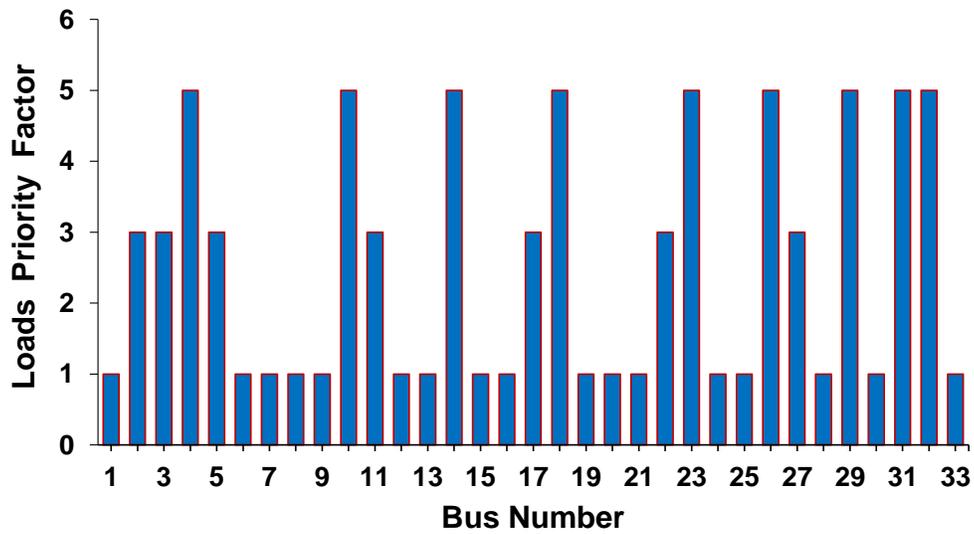

Fig 3. Load priority factor for each bus

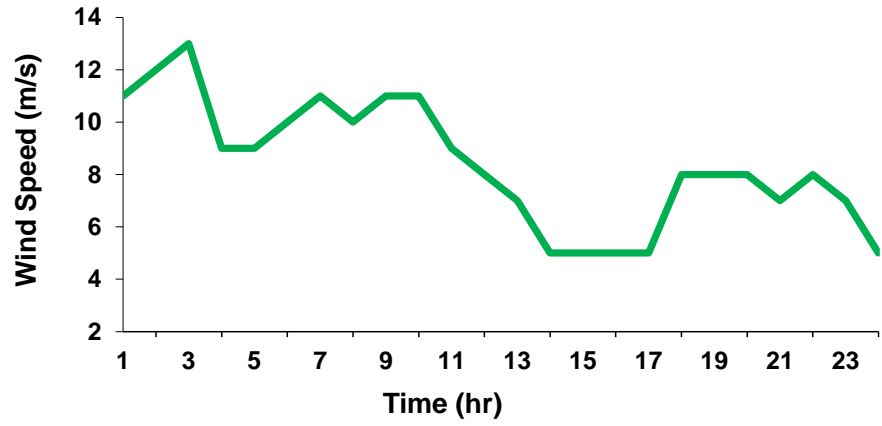

Fig. 4: Wind speed profile during the day

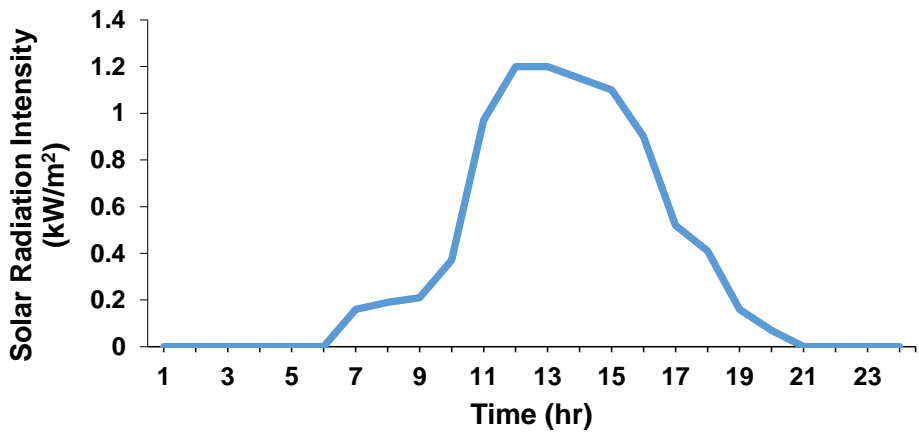

Fig. 5: Solar radiation intensity during the day

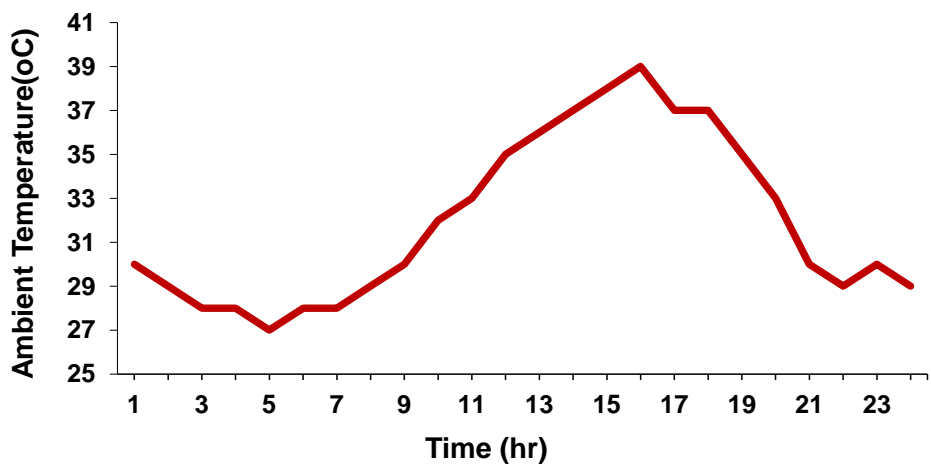

Fig. 6: Ambient temperature during the day

## 3.2. Simulation results

To evaluate performance of the proposed approach, it has been applied to the IEEE 33-bus system for different condition. For this aim, the simulations are implemented for radial and mesh structures with different weighting coefficients. It should be noted that in this network, the disaster causes the outage of line 1-2 that brings about disconnection of distribution system from the main substation. This outage is accounted as a severe fault [1]. Table 3 presents the results for three cases including: one integrated island, and two and three separated islands. It should be noted that in all cases, the configuration of network in normal operation ($c_1$) is always radial, and for the emergency operation, both of radial and mesh structures are examined. All cases have been carried out for 5 states of weightings as shown in Table 3.

Table 2: Parameters of network equipment

| Element | Parameter | Value |
|---|---|---|
| WT | $P_{rated}$ | 200 kW |
| | $V_{c_{in}}$ | 3 m/s |
| | $V_{c_{out}}$ | 25 m/s |
| | $V_{rated}$ | 12 m/s |
| PV | $V_{MPP}$ | 28.66 V |
| | $I_{MPP}$ | 7.76 A |
| | $V_{OC}^{PV}$ | 36.96 V |
| | $I_{SC}^{PV}$ | 8.38 A |
| | $N_{OT}$ | 43 °C |
| | $K_I$ | 0.00545 A/°C |
| | $K_V$ | 0.1278 V/°C |
| | $N_{PV}$ | 972 |
| ESS | $SOC^{Max}$ | 1.2 MWh |
| | $P_{Max}^{ESS}$ | 0.2 MW |
| | $SOC_{i,c,t_0}$ | $0.4 SOC^{Max}$ |
| DG | $S_{Max}^{DG}$ | 500 kW |
| | $PF_c \quad ;\forall c \in c_1$ | 0.9 |
| | $PF_c \quad ;\forall c \in c_2$ | 0.8 |
| Network | $I_\ell^{Max}$ | 250 A |
| | Location of PV units | Buses 10 and 22 |
| | Location of WT units | Buses 3 and 30 |
| | Location of ESS units | Buses 3, 10, 22, and 30 |
| | $V_{i,c,t} = 1 \quad ;\forall c \in c_1, i \in \Omega_S$ | |
| | $0.95 \leq V_{i,c,t} \leq 1.05 \quad ;\forall c \in c_1$ | |
| | $0.9 \leq V_{i,c,t} \leq 1.1 \quad ;\forall c \in c_2$ | |

Table 3: Results of simulation for different number of islands and weightings

| Weighting factors | | $OF_1+OF_2$ | | | | | |
|---|---|---|---|---|---|---|---|
| | | One integrated island | | Two separate islands | | Three separate islands | |
| $w_1$ | $w_2$ | Radial | Mesh | Radial | Mesh | Radial | Mesh |
| 0 | 4 | 40.179 | 40.001 | 38.360 | 38.360 | 53.101 | 49.883 |
| 1 | 3 | 29.189 | 29.173 | 29.375 | 29.375 | 29.574 | 29.573 |
| 2 | 2 | 29.196 | 29.174 | 29.379 | 29.379 | 31.042 | 31.041 |
| 3 | 1 | 29.203 | 29.177 | 29.379 | 29.379 | 31.042 | 31.041 |
| 4 | 0 | 125.094 | 124.986 | 123.105 | 122.912 | 115.479 | 114.549 |

As seen, for the first and fifth weighting factors, i.e., (0, 4) and (4, 0), as the optimization only regards one of the objective functions, the total value ($OF_1+OF_2$) is significantly large. For other weightings, as $w_1$ is increased ($w_2$ is decreased), the total value worsens for both of radial and mesh structures. It can be also observed that as the number of islands is increased, the results of radial and mesh structures are become close together. This can be due to the fact that for forming more islands, more lines must be disconnected, and in each island, there will be fewer sectionalizer-equipped lines to be opened or closed. From Table 3, it is revealed that among three cases, the results of one integrated island for the mesh structure, when $w_1 = 1$ and $w_2 = 3$, is the best among different scenarios. This stems from the fact that in one island case, the number of lines in mesh configuration is more than that of two or three island cases such that the optimization has more flexibility to manage power loss and load shedding. It should be regarded that in this paper, as the emergency condition arises from the outage of the main substation, the remaining part of the network is sound, and hence, the optimization tends to maintain an integrated island. Under the condition that more lines are affected by the natural disaster, off course, configurations with two or more islands may be the optimal state. The other issue regarding the results is that by comparing radial and mesh structures, the mesh one shows better performance than the radial configuration as the greater number of lines prepares more freedom for the optimization to minimize power loss and load shedding. It should be noted again that this table gives only the results of emergency mode (i.e. the mode $c_2$), and for the normal mode (i.e. the mode $c_1$), the configuration of network is always radial.

Regarding the above discussion, the schematic view of the best configurations for the network in normal and emergency condition have been depicted in Figs. 7-9. Fig. 7 is for the normal operation, and Figs. 8 and 9 show the configuration in emergency operation with radial and mesh structures, respectively. It should be noted that for the one, two and three island cases, the layout of normal mode may be different for radial and mesh structures of the emergency mode. In one integrated island, these two layouts are the same, but for the other cases, they are different which will be presented subsequently.

As the results are too many, in this part, only those of one integrated island with mesh structure (the best one of Table 3) are presented in detail. Tables 4 and 5 report the objective functions for different weighting factors in radial and mesh structures. As seen from Table 5, the resiliency index ($RI$), which shows the restored loads during the emergency condition, has the highest value for $w_1 = 1$, and $w_2 = 3$. The reason that the power loss is almost the same in second, third, and fourth weighting factors is due to reaching to the minimum possible value.

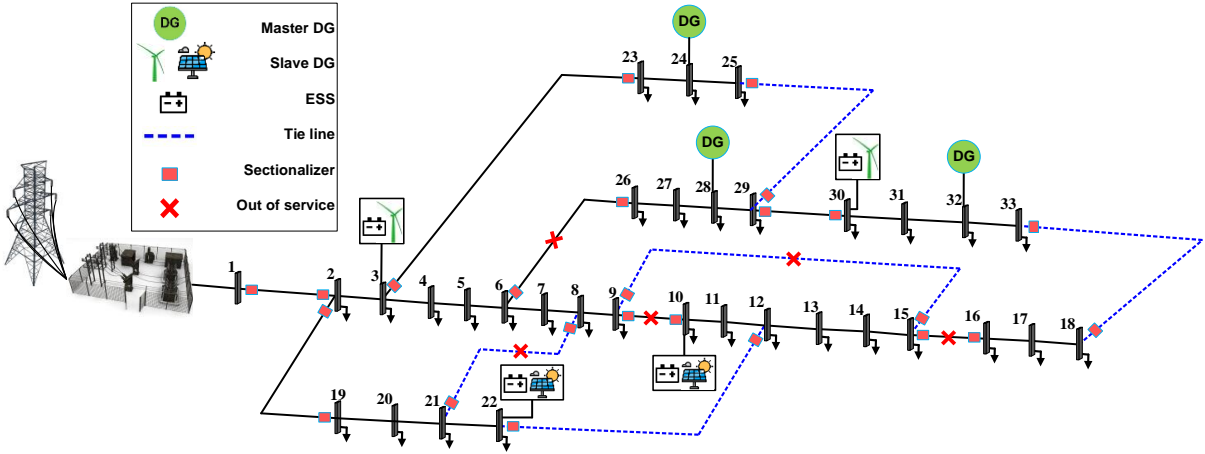

Fig. 7: Schematic view of the system in mode $c_1$ for one island case

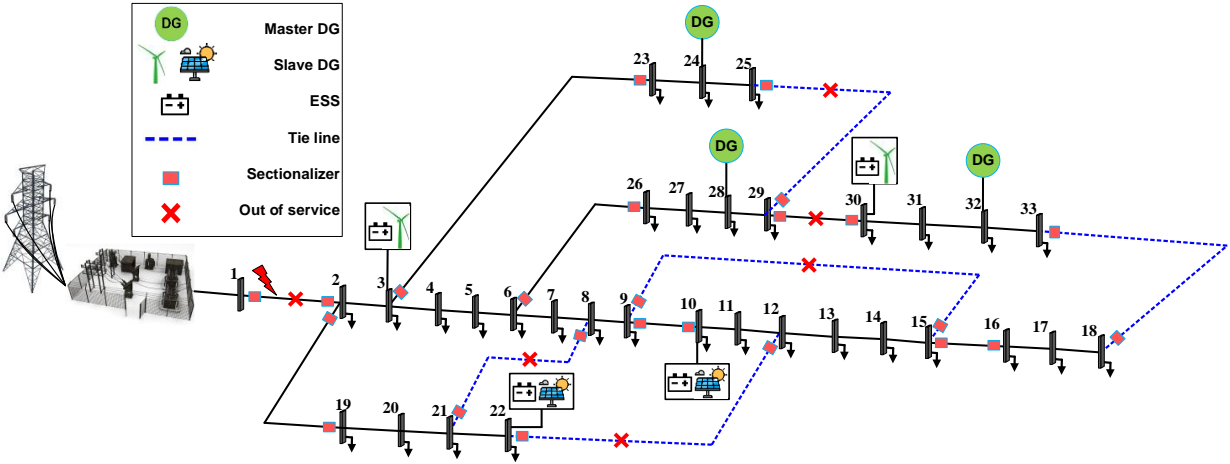

Fig. 8: Schematic view of one island case with radial structure in mode $c_2$

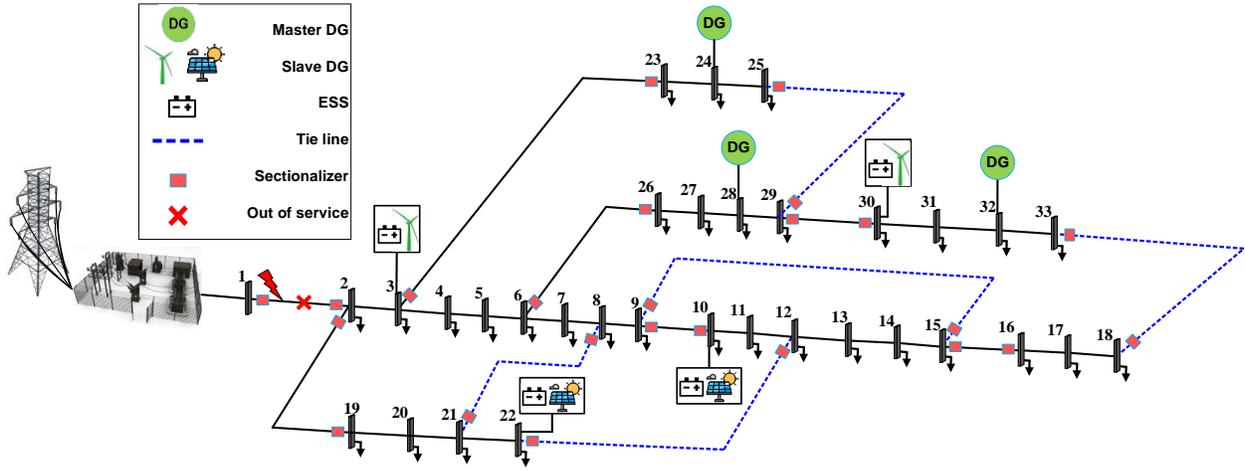

Fig. 9: Schematic view of one island case with mesh structure in mode $c_2$

Table 4: Results of one integrated island with radial structure

| $w_1$ | $w_2$ | $OF_1(MWh)$ | $OF_2$ | $OF$ | $P_c^{Lsh_{total}}(MWh)$ | $RI$ | $OF_1^{optimum}(MWh)$ | $OF_2^{optimum}$ |
|---|---|---|---|---|---|---|---|---|
| 0 | 4 | 11.487 | 28.692 | 1.001 | 28.692 | 57.842 | 0.467 | 28.692 |
| 1 | 3 | 0.467 | 28.721 | 1.001 | 28.721 | 57.800 | 0.467 | 28.692 |
| 2 | 2 | 0.467 | 28.729 | 1.001 | 28.729 | 57.788 | 0.467 | 28.692 |
| 3 | 1 | 0.467 | 28.735 | 1.000 | 28.735 | 57.779 | 0.467 | 28.692 |
| 4 | 0 | 0.467 | 124.627 | 1.000 | 53.472 | 21.432 | 0.467 | 28.692 |

Table 5: Results of one integrated island with mesh structure

| $w_1$ | $w_2$ | $OF_1(MWh)$ | $OF_2$ | $OF$ | $P_c^{Lsh_{total}}(MWh)$ | $RI$ | $OF_1^{optimum}(MWh)$ | $OF_2^{optimum}$ |
|---|---|---|---|---|---|---|---|---|
| 0 | 4 | 11.311 | 28.690 | 0.999 | 28.690 | 57.845 | 0.467 | 28.690 |
| 1 | 3 | 0.467 | 28.706 | 1.000 | 28.706 | 57.822 | 0.467 | 28.690 |
| 2 | 2 | 0.467 | 28.707 | 1.000 | 28.707 | 57.820 | 0.467 | 28.690 |
| 3 | 1 | 0.467 | 28.710 | 1.000 | 28.710 | 57.816 | 0.467 | 28.690 |
| 4 | 0 | 0.467 | 124.519 | 1.000 | 53.364 | 21.591 | 0.467 | 28.690 |

In Fig. 10, the output power of the wind turbines in different buses and hours in modes $c_1$ and $c_2$ are expressed for the one island mesh configuration. It is observed that in $c_1$ mode, the value of scheduled power is as much as the network needs, while in $c_2$ mode, due to the isolation of the network from the main network and shortage of generating resources, all the capacity of wind turbines in the network is used. Fig. 11 shows the generated power of PV units in the mentioned condition. It is seen that the whole available power of photovoltaic units is scheduled at all hours in both modes of $c_1$ and $c_2$.

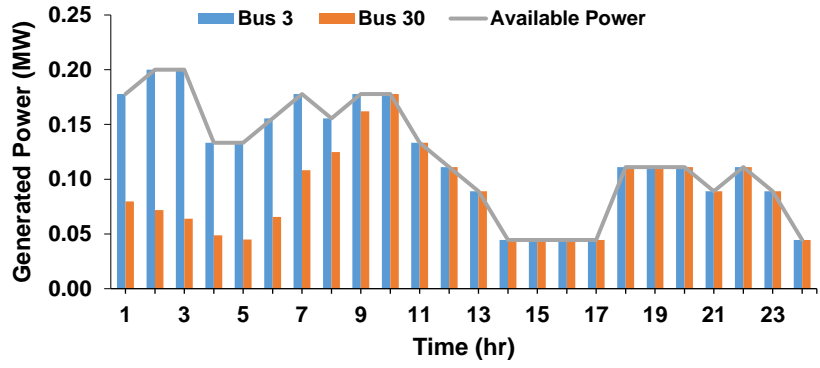

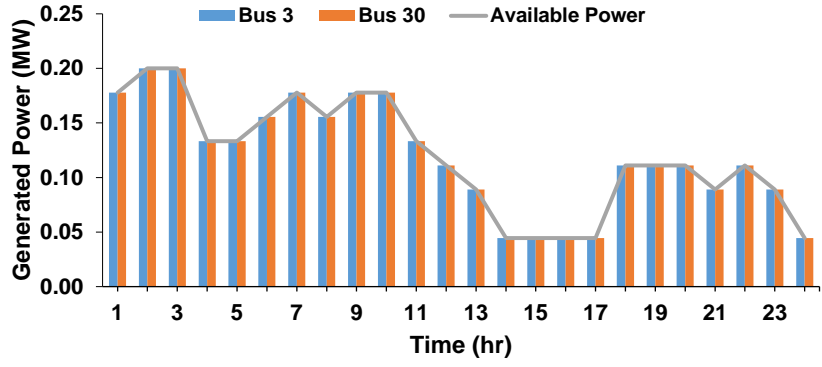

Fig. 10: Active power of wind turbines, (a): mode $c_1$, (b) mode $c_2$

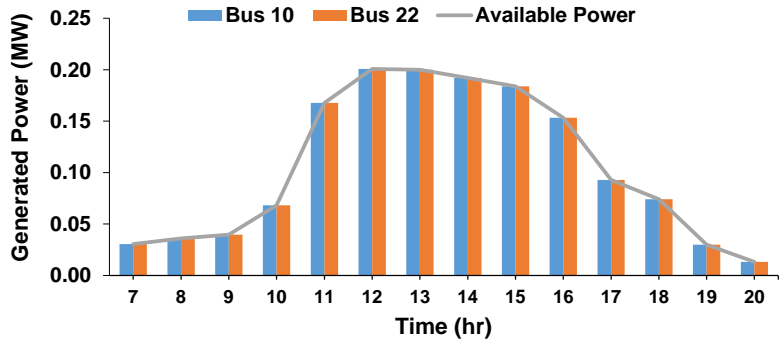

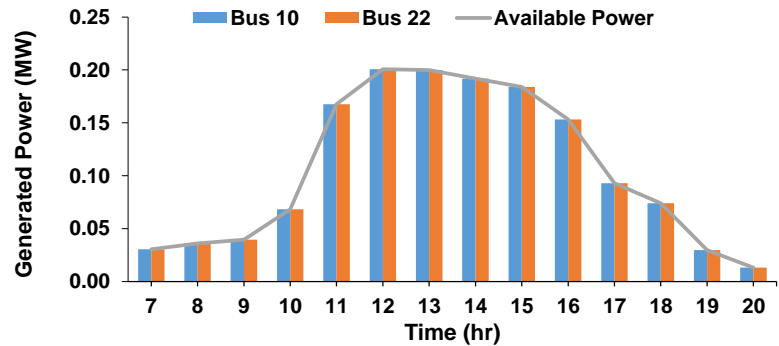

Fig. 11: Active power of PVs, (a): mode $c_1$, (b) mode $c_2$

In Fig. 12, the energy level of ESS units in different buses and hours in modes $c_1$ and $c_2$ are shown for the one island mesh configuration. It is seen that the ESS units 3 and 30 in mode $c_1$ discharge their initial stored energy such that the network receives less power from the upward grid in order to reduce the power loss. They are charged during hours 12 to 20, and again discharged from hour 20 to 24. About ESS units installed at buses 10 and 22, their stored energy is discharged during hours 10 to 16, and the charging period is happened at hours 17 to 23.

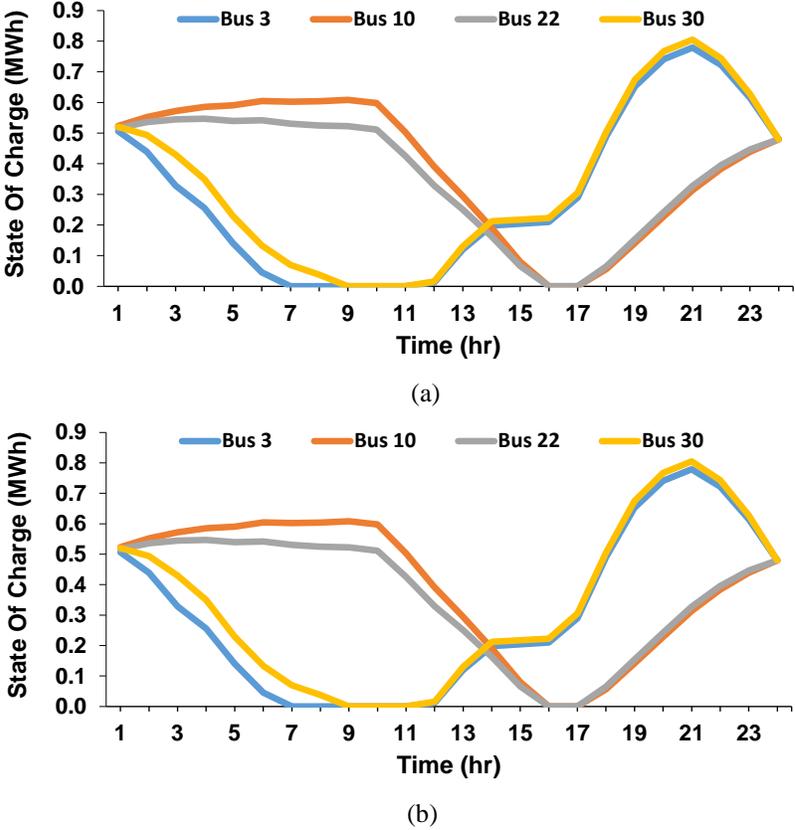

Fig. 12: State-of-charge for ESS units, (a): mode $c_1$, (b) mode $c_2$

The active and reactive power injected from the upward grid into distribution system in mode $c_1$ is shown in Fig. 13. Due to the load increase in peak hours, the received power in these periods are higher. It should be noticed that this injected power is zero in mode $c_2$ as the main substation is out of service due to the natural disaster. In Figs. 14-16, active and reactive powers generated by synchronous (master) DG units have been shown for different modes of one island case considering the mesh structure. The figures state that the units operate with their maximum active power capacity by regarding reactive power generation to assist in improving voltage profile and reducing the power losses. In mode $c_2$, it is observed that there is a reduction of active power generation along with increase of reactive power generation at hour 18. This is due to the need of network for generating more reactive power for compensating high voltage drop at this hour.

To see the voltage condition of the network, the voltage profile at the peak hour of 18 is depicted in Fig. 17 for the two modes. In mode $c_1$, bus 16 has the lowest voltage magnitude which is about 0.97 p.u. Although this profile is for the peak hour, it is seen that the voltage of all buses are within the acceptable range. Also, in mode $c_2$, the voltages are in the permissible bound. In Fig. 18, the value of total load shedding for active and reactive powers in mode $c_2$ has been illustrated. As the distribution system is isolated from the upward grid, and regarding limitation of available resources, there will be more load shedding at the peak hours. Off course, the optimization has tried to recover loads as much as the network constraints are satisfied.

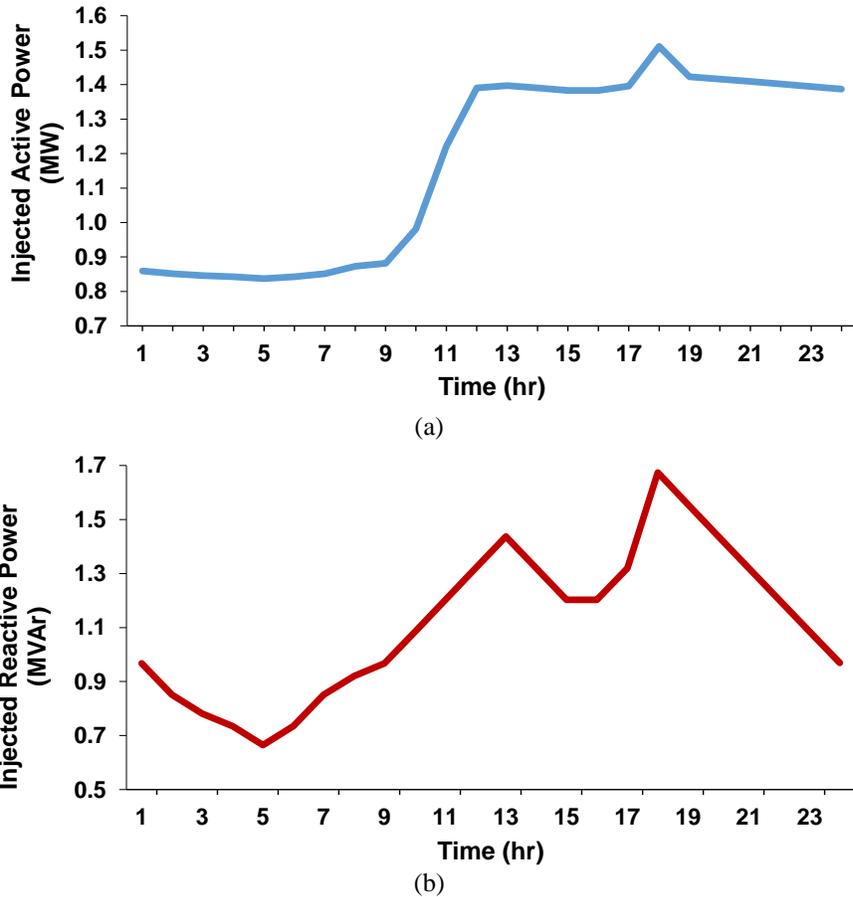

Fig. 13: Active and reactive power injected from upstream network in mode $c_1$, (a): active power, (b): reactive power

As mentioned earlier, the proposed model in this paper for simultaneous resiliency improvement in emergency condition and optimal operation in normal condition has the ability of forming any number of islands in mode $c_2$. In the previous part, the objective function value for different number of islands where reported in Table 6. Also, because of huge volume of output results, only the detailed results of one integrated island were described. In this section, the layout of three-island case is shown for instance. Figs. 19 to 22 illustrate these configurations for radial and mesh structures in both modes of $c_1$ and $c_2$.

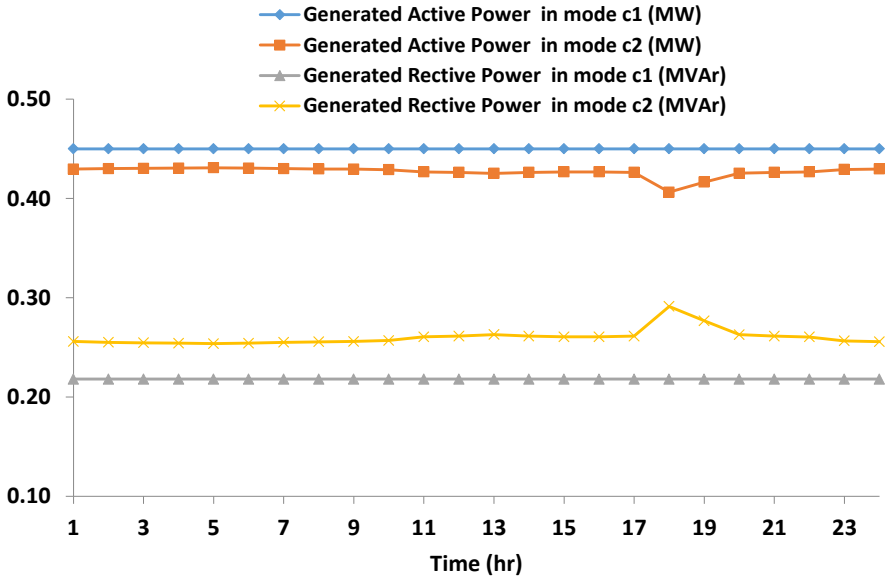

Fig. 14: Generated active and reactive powers of synchronous DG located @ bus 24 in different modes

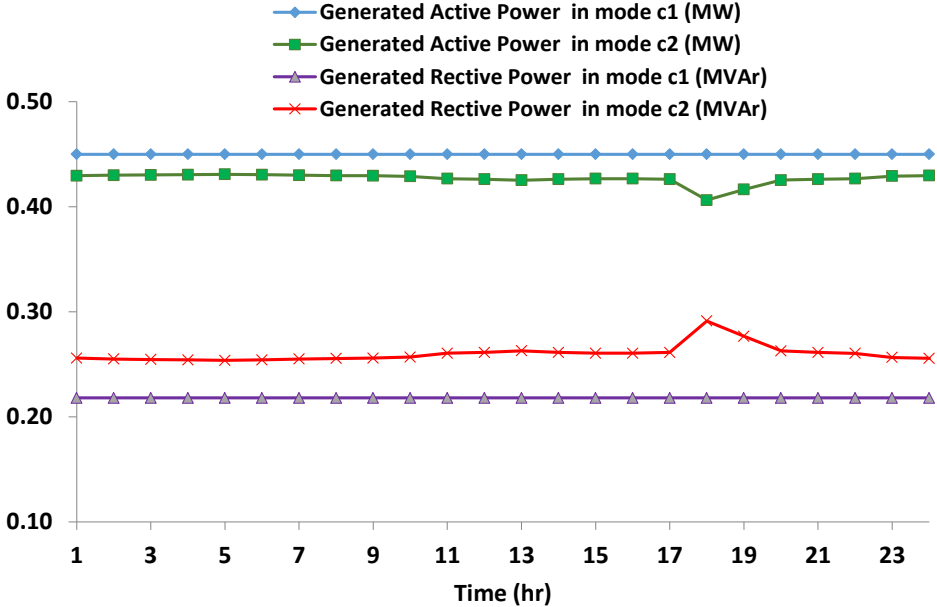

Fig. 15: Generated active and reactive powers of synchronous DG located @ bus 28 in different modes

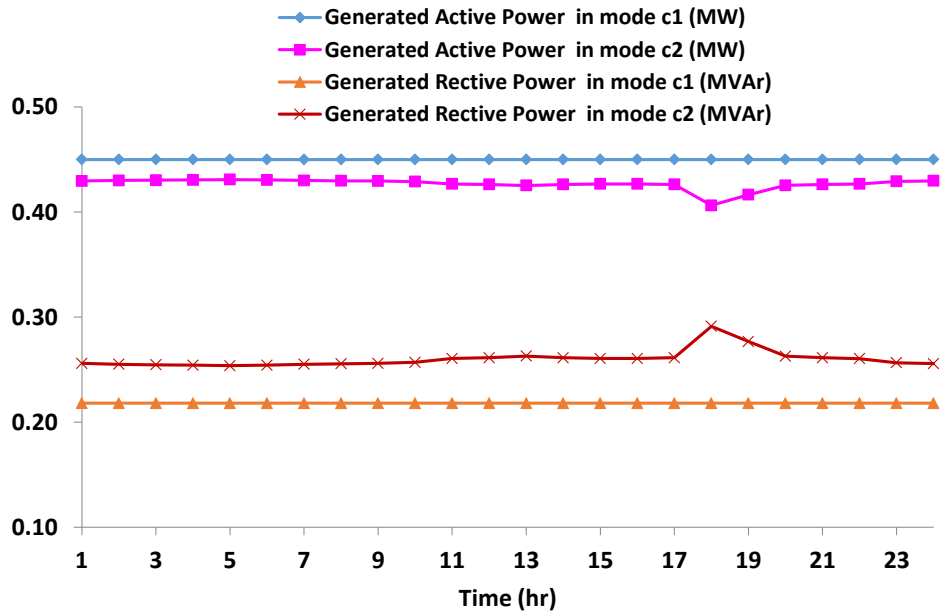

Fig. 16: Generated active and reactive powers of synchronous DG located @ bus 32 in different modes

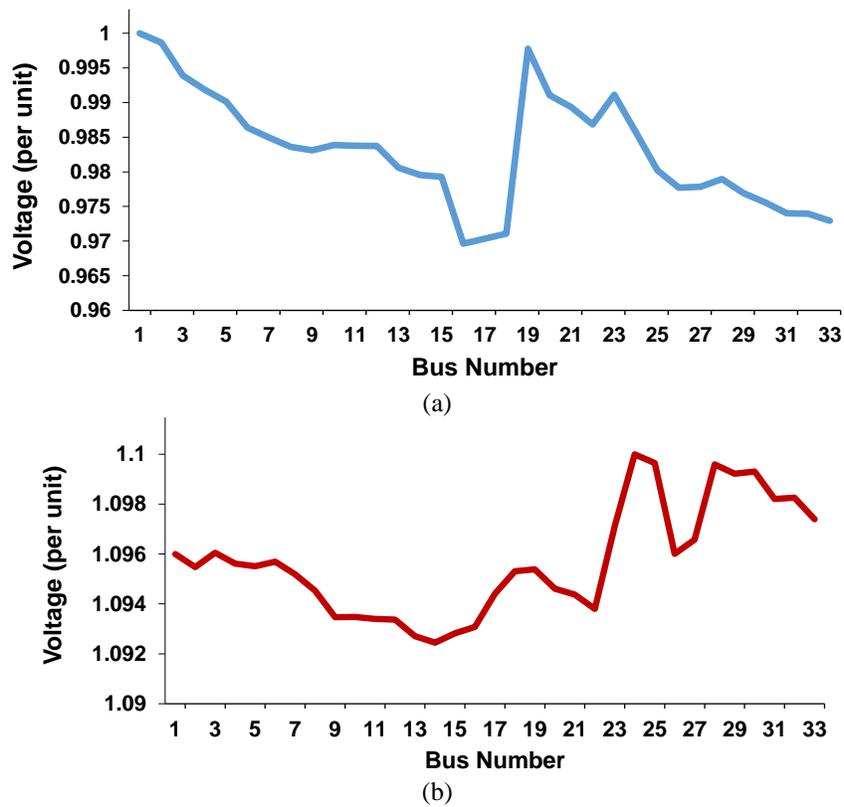

Fig. 17: Voltage profile of the network at peak hour in mode $c_1$ and $c_2$, (a): mode $c_1$, (b): mode $c_2$

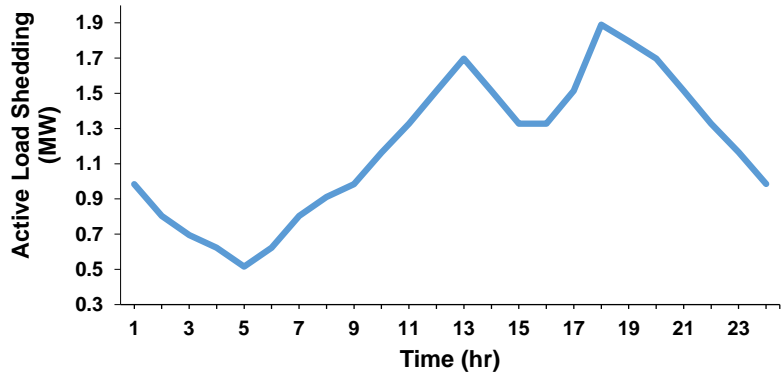

(a)

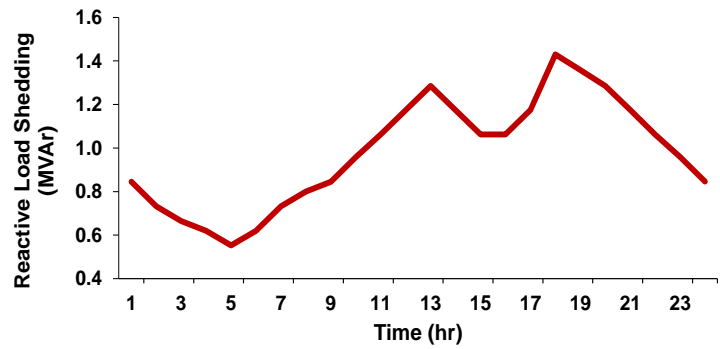

(b)

Fig. 18: The total active and reactive load shedding in mode $c_2$, (a): active power, (b): reactive power

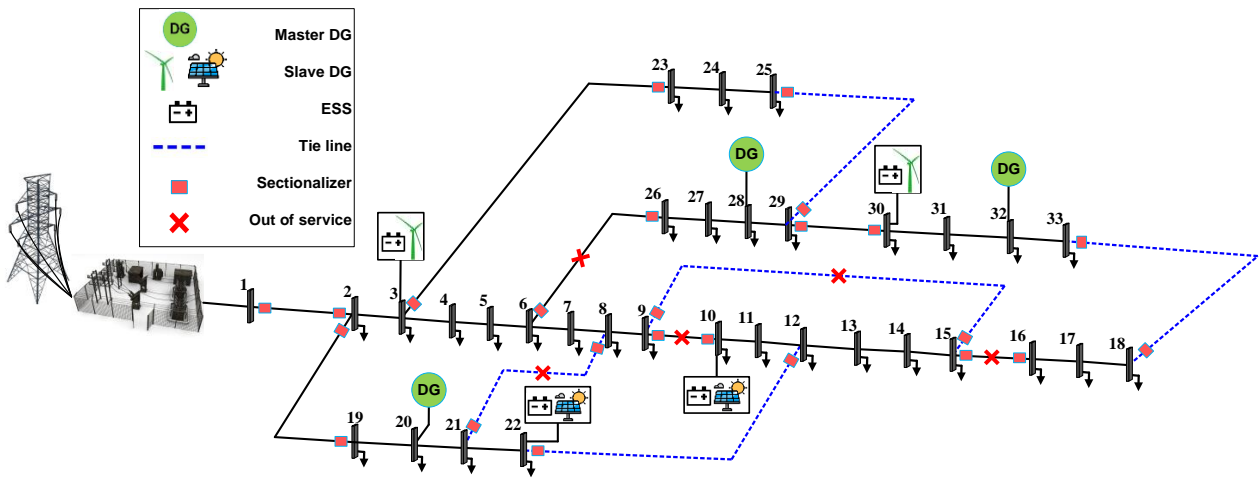

Fig. 19: Schematic view of three island case with radial structure in mode $c_1$

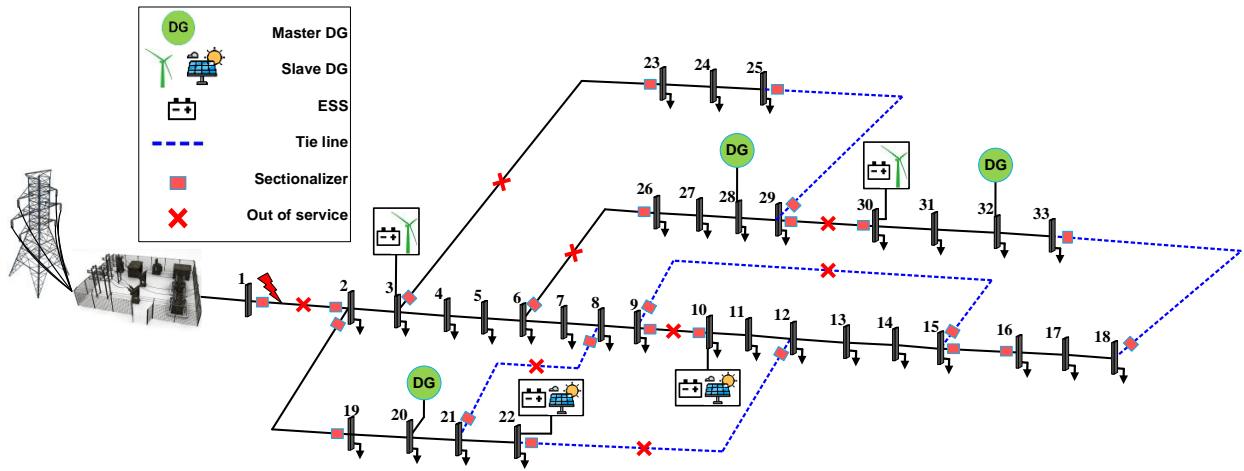

Fig. 20: Schematic view of three island case with radial structure in mode $c_2$

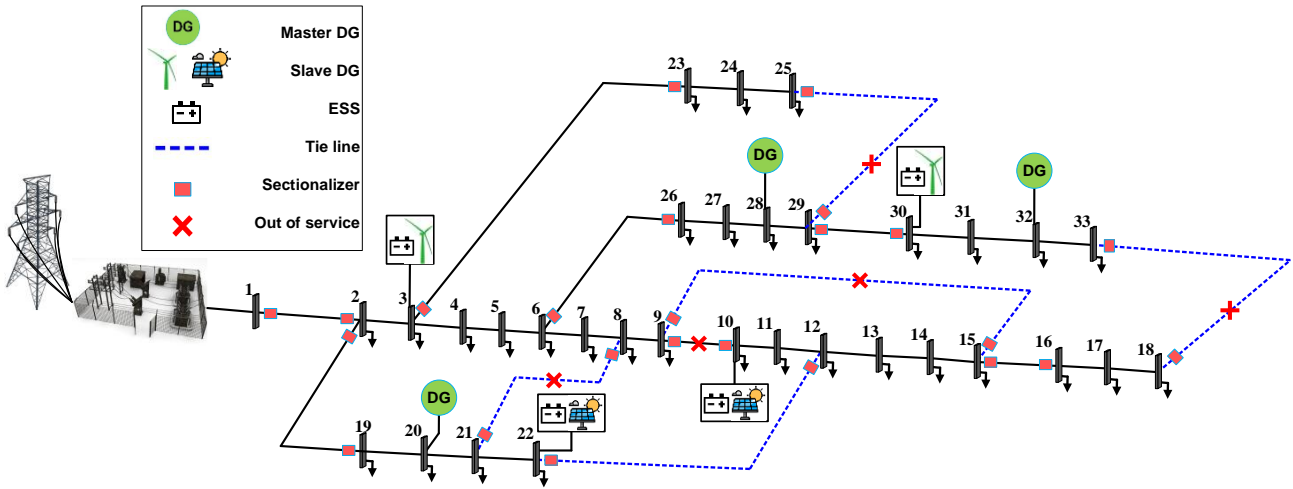

Fig. 21: Schematic view of three island case with mesh structure in mode $c_1$

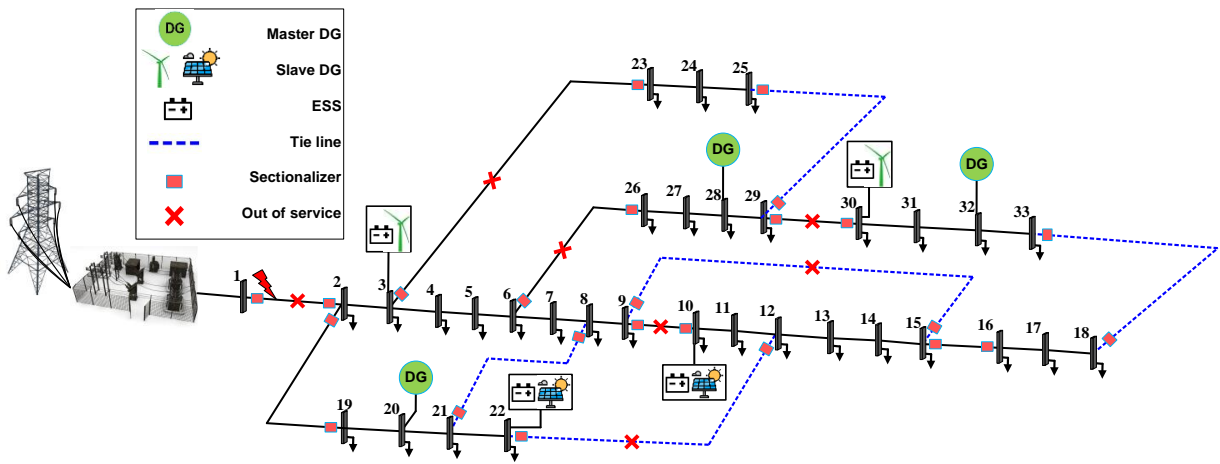

Fig. 22: Schematic view of three island case with mesh structure in mode $c_2$

## 4. Conclusion

In this paper, a new model was proposed for optimal operation of sustainable distribution network considering both normal and emergency conditions. The presented methodology employs network reconfiguration, micro-grid formation, and distributed energy resources allocation to enhance resiliency of system in critical conditions and minimize the power loss during normal operating condition. Based on new formulations, optimal micro-grids are formed as radial and mesh structures to restore maximum load after severe outages. The AC power flow equations based on LFB method along with all other relations have been linearized to make a convex model as MIQCP. By applying the conducted approach on the IEEE 33-bus system, its efficiency has been investigated through different studies. The simulations verify optimal operation of network in both normal and emergency conditions along with satisfying all the problem constraints. The results of radial and mesh structure in emergency conditions, verified better performance of the mesh one in terms of resiliency index. The developed model can be also applied when more places of the network are affected by the natural event such that the model can constitute any number of MGs in an optimal manner. It can be also applied to any desired network with any scale and with any number of disaster-affected areas. For the future researches, it is recommended to reinforce the network equipment using installation and expansion of lines and substations. Also, the uncertainty of renewable power generations will be considered in future works. Forecasting of possible failed lines caused by natural disasters and network planning based on this forecast can be accounted as the other research direction.